\documentstyle[twocolumn,epsfig,color]{mn}
\newcommand{\mincir}{\raise
-2.truept\hbox{\rlap{\hbox{$\sim$}}\raise5.truept
\hbox{$<$}\ }}
\newcommand{\magcir}{\raise
-2.truept\hbox{\rlap{\hbox{$\sim$}}\raise5.truept
\hbox{$>$}\ }}
\newcommand{\minmag}{\raise-2.truept\hbox{\rlap{\hbox{$<$}}\raise
6.truept\hbox{$>$}\ }}
\newcommand{\be}{\begin{equation}}
\newcommand{\ee}{\end{equation}}
\newcommand{\ba}{\begin{eqnarray}}
\newcommand{\ea}{\end{eqnarray}}
\newcommand{\brr}{\begin{array}}

\newcommand{\err}{\end{array}}
\newcommand{\bc}{\begin{center}}
\newcommand{\ec}{\end{center}}

\newcommand{\mnras}{MNRAS}
\newcommand{\apj}{ApJ}
\newcommand{\apjl}{ApJL}
\newcommand{\apjs}{ApJS}
\newcommand{\aj}{AJ}
\newcommand{\aap}{A\&A}
\newcommand{\nat}{Nature}

\title[Composite star formation histories of ETGs]{Composite star formation histories of early-type galaxies from minor mergers: prospects for WFC3}

\author[S.~Peirani et al.]
 {S. Peirani$^{1,2}$\thanks{E-mail: peirani@iap.fr},
  R.M. Crockett$^{2}$, S. Geen$^{2}$,  S. Khochfar$^{3}$, S. Kaviraj$^{2,4}$ \& 
 J. Silk$^{2}$\\
$^{1}$ Institut d'Astrophysique de Paris, 98 bis Bd Arago, 75014 Paris, France -\\
 Unit\'e mixte de recherche 7095 CNRS - Universit\'e Pierre et Marie Curie. \\
$^{2}$ Department of Physics, University of Oxford, Denys Wilkinson Building,
 Keble Road, Oxford OX1 3RH, UK\\
$^{3}$  Max Planck Institut f\"ur extraterrestrische Physik, p.o. box 1312, D-85478 Garching, Germany\\
$^{4}$ Blackett Laboratory,
Imperial College London,
South Kensington Campus,
London SW7 2AZ\\
}

\begin{document}

\maketitle

\begin{abstract}
The star formation history of nearby early-type galaxies is investigated
via numerical modelling. Idealized hydrodynamical N-body simulations
with a star formation prescription are used to study the minor merger
process $( 1/10 \leq M_1/M_2 \leq 1/4; M_1 \leq M_2)$
between a giant galaxy (host) and a less massive spiral galaxy (satellite)
with reasonable assumptions for the ages and metallicities of the merger progenitors.
We find that the evolution of the star formation rate is extended over several
dynamical times and shows peaks which correspond to  pericentre passages of the satellite. 
The newly formed stars are mainly located in the
central part of the satellite remnant while the older stars of the initial disk
are deposited at larger radii in shell-like structures. After the final plunge of
the satellite, star formation in the central part of the remnant can continue for
several Gyrs depending on the star formation efficiency.
 Although the mass fraction in new  stars is small, we find that the half-mass radius
differs  from the half-light radius in the V and H bands. 
Moreover synthetic 2D images in  J, H, NUV, H$_\beta$ and V bands, using
the characteristic filters
of the Wide Field Camera 3 (WFC3) on the Hubble Space Telescope (HST), reveal that
residual star formation induced by gas-rich minor mergers can be clearly observed
during and after the final plunge, especially in the NUV band, for interacting systems
at ($z \leq 0.023$) over  moderate numbers of orbits
($\sim$ 2 orbits correspond to  typical exposure times of $\sim$ 3600 sec).
This suggests that  WFC3 has the potential to resolve these substructures,
characterize plausible  past merger episodes,  and give clues to the formation
of early-type galaxies.

\end{abstract}


\section{Introduction}

The formation and evolution of early-type galaxies (hereinafter ETGs)
remains an outstanding puzzle in contemporary astrophysics
(see Cimatti 2009 for a recent review). According to earlier studies, ETGs seem
to form a class of objects with simple and well defined properties:
they tend to have a smooth morphology, an old stellar population,
a red optical colour and are free of cold gas and ongoing star formation
(Searle, Sargent \& Bagnulo 1973; Larson 1975). However, this classic
picture is challenged by recent observations.
For instance, the presence of cold gas and dust as well as
evidence of recent star formation has been detected in several
systems (see, for instance, Goudfrooij et al. 1994; Macchetto et al.
1996; Trager et al. 2000; Yi et al. 2005; Morganti et al. 2006;
Kaviraj et al. 2007; Sarzi et al. 2008, Clemens et al. 2009b), and the associated
star formation rate can even be comparable to that in normal spiral galaxies
although these events are rather rare (Fukugita et al. 2007).
Moreover, recent studies based on GALEX (Galaxy
Evolution Explorer) near-UV photometry of a large sample of ETGs found
that 30\% of them experience ongoing or recent star formation (Schawinski et
al. 2007; Kaviraj et al. 2007), and a similar trend has been
obtained from mid-infrared Spitzer data in the Coma and Virgo Clusters
(Bressan et al. 2006; Clemens et al. 2009a)
or from the COSMOS survey (Kaviraj et al. 2010).

Understanding the formation of early-type galaxies, and in particular their
star formation history, is  of crucial importance for setting strong
constraints on models of galaxy formation. As has been widely discussed
in the  literature, there are currently two competing scenarios
for the formation of ETGs. On the one hand there is the {\it monolithic collapse}
model in which galaxies are formed in short, highly efficient
starbursts at high redshift ($z \gg 1$) and evolve purely passively thereafter
(Eggen, Lynden-Bell \& Sandage 1962; Larson 1974; Arimoto \& Yoshii 1987).
On the other hand, the {\it hierarchical structure formation} model
suggests that the most massive galaxies are formed from successive mergers
of smaller entities with accompanying star formation over a cosmological time-scale (Toomre 1977;
Blumenthal et al. 1984; Kauffmann, White, \& Guiderdoni 1993). Although
these models are in good agreement with some aspects of observational data,
they both face some problems. For instance, the monolithic model is
supported by the existence of, and small scatter in, galaxy scaling relations
such as the colour-magnitude relation (Sandage \& Visvanathan 1978), the
fundamental plane (Djorgovski \& Davis 1987; Jorgensen, Franx \& Kjaergaard 1996;
Saglia et al. 1997) and the Mg-$\sigma$ relation (Colless et al. 1999;
Kuntschner et al. 2001),  but cannot easily explain the residual
star formation and cold gas found in ETGs. As far as hierarchical 
models are concerned, although they predict galaxy interaction, they are in
disagreement with  the so-called ``down sizing''  phenomenon. Indeed, observations
of deep surveys ($z\geq 1-2$), such as the Las Campanas Infrared
Survey,  HST Deep Field North and Gemini Deep Deep Survey (GDDS)
have revealed an excess of massive early-type galaxies undergoing
``top-down'' assembly with high inferred specific star formation
rates relative to predictions of the hierarchical scenario
(Kodama et al. 2004; Glazebrook et al. 2004; Cimatti, Daddi \& Renzini 2006).

It is now well known that ETGs have considerable substructure 
(e.g. from SAURON and GALEX) which is interpreted as  a result
of mergers in the past several gigayears.
 To study this plausible process,  we have
compared the ultra-violet (UV) colours of nearby ($0.05 \leq z \leq 0.06$)
early-type galaxies with synthetic photometry derived from numerical
simulations of minor mergers, with reasonable assumption for the ages,
metallicities and dust properties of the merger progenitors
 (Kaviraj et al. 2009, hereafter K09).
We found that the large scatter in the ultra-violet colours of intermediate
mass early-type galaxies in the local universe and the inferred
low-level recent star formation in this objects can be reproduced by
minor mergers in the standard $\Lambda$CDM cosmology.
In the present work, our aim is to study the evolution of the internal structure
of these objects using the same methodology but with higher mass resolution,
in order to help understand in more detail the observational signatures
of satellite minor merger events with different mass ratios, gas-fractions
and orbital configurations. This work is also motivated by the recent
installation on the Hubble Space
Telescope (HST) of NASA's Wide Field Camera 3
(WFC3\footnote{http://www.stsci.edu/hst/wfc3})
whose optical design provides a large field of view
and high sensitivity over a broad wavelength range,  excellent
spatial resolution and a stable and accurate photometric performance.
It features two independent imaging cameras, a UV/optical channel
(UVIS) and a near-infrared channel
(IR). Both channels are equipped with a broad selection of spectral filters
and some of these have been considered in our study in order to
identify disrupted satellites and trace the history of the merger process.

This paper is organized as follows: in section 2 we summarize the
numerical methodology that we have developed to produce WFC3 synthetic images;
in section 3 we present results on the star formation history
of nearby ETGs,  while the last section presents our conclusions.

\section{Numerical methodology}

\subsection{Simulations}

The numerical methodology used in the present paper is described
in detail in K09 where we refer the reader for more information.
For the sake of clarity, we summarize the main steps below.

Our aim is to study minor mergers  between an elliptical galaxy (E)
and a satellite (spiral galaxy) using idealised N-body+SPH
simulations.  Both  galaxies are created following 
Springel, Di Matteo \& Hernquist (2005). First,
the elliptical is modelled using a spherical dark matter (DM) halo
and a stellar component, with a total mass of $10^{12} M_{\odot}$.
Stars contribute 4\% of this value in agreement with recent observational
studies (see for instance, Jiang \& Kochanek 2007; McGaugh et al. 2009).
A Hernquist
profile reproduces the de Vaucouleurs $R^{1/4}$ surface
brightness profiles of typical elliptical galaxies. The effective
radius of the projected brightness is $r_e=4.29$ kpc.
As far as the satellite is concerned, we used a compound galaxy model which
consists of a spherical DM halo and a rotationally supported disk
of gas and star (but no bulge) with independent parameters describing each of
the structural component.
 The mass of the disk represents 5\% of the total
mass of the satellite and the gas mass fraction in the disk is fixed to
a constant value 20\% (of the total mass of the disk), consistent with
observed values from the SDSS (Kannappan 2004). In all simulations,
satellites are put on prograde or retrograde parabolic orbits
(Khochfar \& Burkert, 2006),
with varying pericentric distance $r_{p}$ and initial separations of $r_{ini}=100$ kpc.

The simulations are performed using the publicly available code GADGET2 
(Springel 2005) with added prescriptions for cooling, star
formation and feedback from Type Ia and II supernovae (SN).
It is worth mentioning that gas particles with
$T< 2\times 10^4 K$, number density $n > 0.1\, cm^{-3}$,
overdensity $\Delta \rho_{gas}> 100$ and ${\bf \nabla . \upsilon}
<0$ form stars according to the standard star formation prescription:
$d\rho_*/dt = c_* \rho_{gas}/t_{dyn}$, where $\rho_*$ refers to
the stellar density, $t_{dyn}$ is the local dynamical
timescale of the gas and $c_*$ is the star formation efficiency.

In the present study, we use 10 times more particles than in K09
in order to more accurately study  the evolution of the spatial 
distribution of stars. 
Consequently, $\sim 4,000,000$ particles are used for each experiment in which 
the particle mass are $M_{DM}=3.03\times10^5 M_\odot$,
$M_{gas}=M_{disk}=4.5\times10^4 M_\odot$ and $M_{E}=1.35\times10^5  M_\odot$
for the dark matter particles (DM), gas particles, star particles in the disk and
star particles in the elliptical galaxy (E) respectively. 
The gravitational softening lengths ($\epsilon$)
used in each simulation are ({\it physical})
$\epsilon_{DM} = \epsilon_{E}=0.1$ kpc  and $\epsilon_{gas} =  \epsilon_{disk}=0.2$ kpc.
 We have also checked that increasing the resolution of the simulations
 as respect to K09
does not affect the star formation history
for any given experiment which suggests that 
 the conclusions presented in
this work are robust. For instance, 
 an example (experiment $B_2$) 
is exibited in  Fig. 3 of K09.

We have considered several realizations whose relevant free parameters
(e.g. initial orbital configuration and star formation efficiency) are
summarized in Table \ref{table1}. Our investigation mainly focuses on
minor satellite mergers ( $1/10 \leq$ mass ratio $\leq 1/4$) but we have also
considered one major merger case (run $D_1$) with which to compare the results.

\begin{table}
\begin{center}
\caption{Summary of merger simulations.
} 
\begin{tabular}{|c|c|c|c|c|}
\hline
run$^a$ & mass ratio$^b$ & $c_*^b$ & $r_p^d$ & orbit$^e$ \\
\hline
$A_1$ &  1:10 & 0.01 & 8 & prograde  \\
$A_2$ &  1:10 & 0.05 & 8 & prograde \\
\smallskip
$A_3$ &  1:10 & 0.1  & 8 & prograde \\
$B_1$ &  1:6  & 0.01 & 8 & prograde \\
$B_2$ &  1:6  & 0.05 & 8 & prograde \\
$B_3$ &  1:6  & 0.1  & 8 & prograde \\
$B_4$ &  1:6  & 0.05 & 6 & prograde \\
$B_5$ &  1:6  & 0.05 & 4 & prograde \\
$B_6$ &  1:6  & 0.05 & 2 & prograde \\
$B_7$ &  1:6  & 0.05 & 0 & prograde \\
\smallskip
$B_8$ &  1:6  & 0.05 & 8 & retrograde \\
$C_1$ &  1:4 & 0.01 & 8 & prograde \\
$C_2$ &  1:4 & 0.05 & 8 & prograde \\
\smallskip
$C_3$ &  1:4 & 0.1  & 8 & prograde \\
$D_1$ &  1:1 & 0.05 & 8 & prograde \\
\hline
\end{tabular}
\label{table1}
\end{center}
\vspace{-0.3cm}
{\scriptsize
$^a$Name of run

$^b$Merger mass ratio (host:satellite)

$^c$Star formation efficiency

$^d$Pericentric distance of orbit in kpc

$^e$type of orbit
}
\end{table}

\subsection{WFC3 synthetic images modelling}

This section describes the numerical modelling
that  enables the rendering of
physically realistic telescope views with 
user-definable filter curves for a simulated
instrument. The rendering process is done
using ray tracing of light emitted by the
simulated objects. For such purpose, we have developed and 
used the Alice
 code\footnote{www-astro.physics.ox.ac.uk/$\sim$samgeen/alice/index.html}.


\begin{figure}
\begin{center}
\rotatebox{0}{\includegraphics[width=6cm]{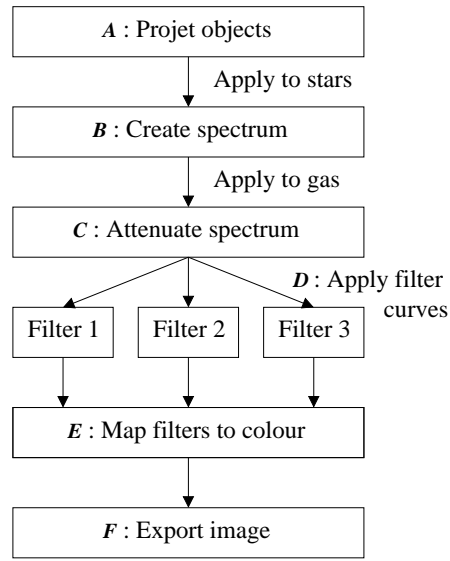}}
\caption{Flow diagram describing the pipeline for the rendering of
telescope views of the objects in the simulation}
 \label{alice}
\end{center}
 \end{figure}

The instrument view render pipeline is
described in Fig. \ref{alice}. 
Firstly, the star and gas particles are projected onto
a 2D array corresponding to the instrument's charge-couple device
(CCD) pixels (A).  This is done using a cosmological distance ladder,
assuming a $\Lambda$CDM cosmology (e.g. Hogg, 1999), with
$H=70$ km s$^{-1}$ Mpc$^{-1}$, $\Omega_m=0.3$ and
$\Omega_{\Lambda}=0.7$ and a flat universe ($k=0$).
Next, we assign a stellar energy distribution (SED) and emission
line strength for various emission lines to each star particle ({\bf B}),
following Bruzual \& Charlot (2003). To do this, we use an
empirically-determined age-metallicity relationship, derived
from Prochaska et al. (2003). The spectra are frequency-shifted
according to the Hubble and peculiar velocity relative to the
simulated observer.
Then the code runs the emission from the star particles and then the
absorption from the gas particles.
Each of a given pixel's spectra and emission
line strengths are summed, and attenuated  ({\bf C}) according to the
gas column density (Calzetti et al. 2000, 2001).
It is worth mentioning that the resolution of the simulations is not enough
to resolve dense molecular gas or dust. Therefore this may underestimate
the total absorption which could affect the  predicted UV emission.
 However, we don't think the
 dust attenuation is very high since
 we showed in K09 that the (integrated) UV colours from this set
of simulations reproduces the observed UV colours of the early-type
galaxy populations extremely well. 
If we were systematically underestimating the dust content then we would
probably see quite a big mismatch between the UV red envelope predicted
by the model simulations and the observed one. 
In the next step, these spectra are passed through a set of transmission filters 
 ({\bf D}). To do this, we linearly interpolate the spectra
calculated with the given transmission curves. Finally, we produce
flux maps corresponding to each transmission filter and emission
line  ({\bf E}). We scale the fluxes to the minimum flux detectable by
 the instrument in each wavelength.

For the purposes of this work, the instrument
parameters were taken from the WFC3
instrument. For instance, we have considered
5 filters: 2 from the near-infrared (NIR) field,   J (F110W)
and H(F160W), and 3 from the UV/optical field, 
 NUV(F225W), the narrow-band H$_\beta$ (F487N)
and V (F555W) (see tables 2 and 3 for the associated wavelength bands).
The UV colours are particularly suitable to detect young stellar populations
inside interacting galaxies (see for instance, Hibbard et al. 2005)
 and remains largely unaffected by the age-metallicity degeneracy
(Worthey 1994)
 while NIR spectral window traces primarily the old stellar populations. 

Synthetic 2D images have been produced assuming luminosity distances of $D=100$ Mpc
and $D=20$ Mpc corresponding
to $z \sim 0.023$ and $z \sim 0.005$ respectively and according the cosmological
parameters considered.  An exposure time,
t = $3600\,s$ was used for each of the broadband filters, while t = $7200\,s$
was used for the narrow-band H$_\beta$ filter.  We have assumed that each  image
is made up of 10 exposures in order to calculate an appropriate readout noise.
The Field of View (FOV) of our synthetic images match those of the WFC3 detectors,
namely: $162\times162$ arcsec for UVIS images and
$133\times133$ arcsec for IR images. All the images have been convolved
with an appropriate HST Point-Spread Function (PSF),
derived   using the {\it TinyTim}\footnote{www.stsci.edu/software/tinytim}
 software package (Krist \& Hook 2004).
Finally, an appropriate sky background, Poisson noise and readout noise 
have been added to each of the filtered images. 
Although generated at the full resolution of the UVIS instrument ($4096\times4096$ pixels),
the UVIS images presented in this paper have been re-binned to a resolution of $1024\times1024$. 

\begin{table}
\begin{center}
\caption{UVIS filters characteristics} 
\begin{tabular}{c|c|c|}
\hline
F225W & F487N & F555W\\
\hline
0.2 - 0.275 $\mu m$& 0.48 - 0.49  $\mu m$ & 0.45 - 0.7  $\mu m$\\
\hline
\end{tabular}
\label{table2}
\end{center}
\end{table}

\begin{table}
\begin{center}
\caption{IR filters characteristics} 
\begin{tabular}{c|c|}
\hline
F110W & F160W \\
\hline
 0.9 - 1.4 $\mu m$&  1.4 - 1.7 $\mu m$\\
\hline
\end{tabular}
\label{table3}
\end{center}
\end{table}

\section{Results}

\subsection{Evolution of the star formation rates}

\begin{figure*}
\rotatebox{0}{\includegraphics[width=17cm]{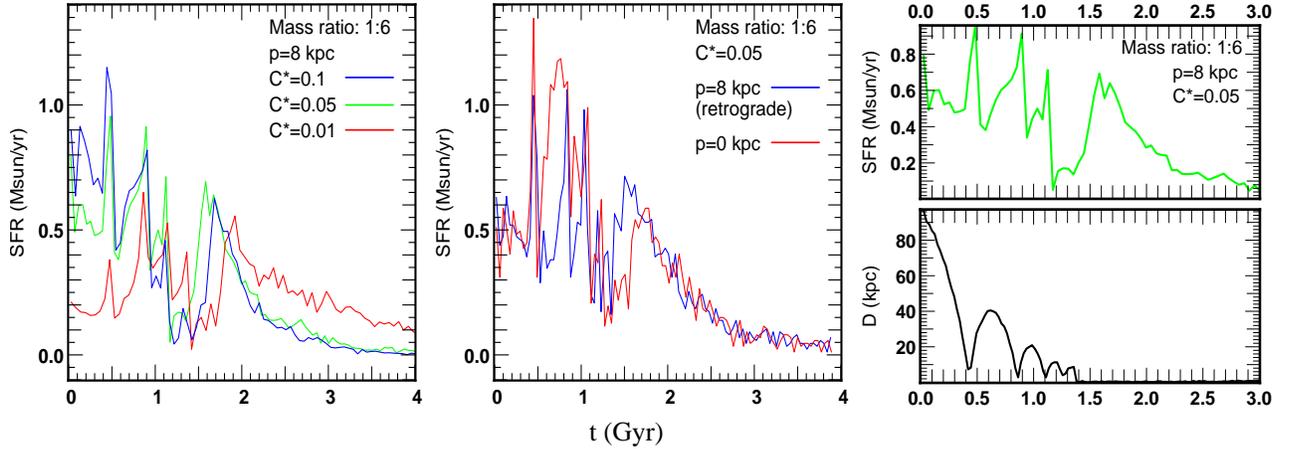}}
\caption{The variations of the star formation rates derived from
1:6 merger mass ratio  scenarios. The left panel shows the SFRs evolution
for a fixed pericentric distance ($r_p=$ 8 kpc)
but with different star formation efficiencies $c_*$ values. The middle panel presents SFRs 
derived from experiments $B_8$ (retrograde orbit) and  $B_7$ 
(edge-on parabolic merger). In the right panel, we have plotted in
parallel the SFR derived from experiment $B_2$ and the evolution of the radial
distance of the most bound part of the satellite.}
 \label{sfr1}
 \end{figure*}

\begin{figure*}
\rotatebox{0}{\includegraphics[width=17cm]{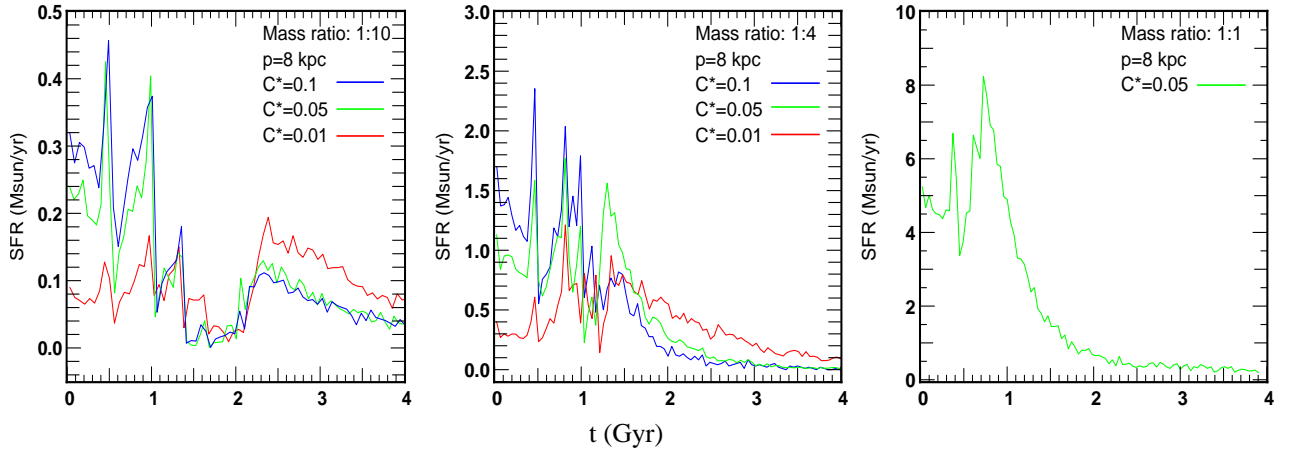}}
\caption{The variations of the star formation rates derived from
1:10 (left panel), 1:4 (middle panel) and 1:1 merger mass ratio 
scenarios.}
 \label{sfr2}
 \end{figure*}

In Figs. {\ref{sfr1}} and {\ref{sfr2}}, we present the time evolution
of the star formation rates (SFRs) derived from  
experiments described in the previous section.                       
In each case, we have let the system evolve until $t=4.0$ Gyr
 from the beginning of the simulation, 
  which corresponds to $\sim 30$
t$_{dyn}$  where t$_{dyn} = (r^3_{vir}/GM_{host}(r_{vir}))^{1/2}$ is the 
dynamical timescale at the virial radius of the host elliptical galaxy.
In our models,  $r_{vir}\approx$ 13.5 kpc and  $M_{host}(r_{vir})\approx
3.12\times 10^{10} M_\odot$ which lead to t$_{dyn}\sim 1.3\times 10^8$ yr.
This value is consistent with observationally determined
dynamical times from SDSS parameters (petroRad, photometric
stellar masses, etc...) that are around 0.1 Gyr for ETGs.
Young (2002) has also
studied interferometric images of CO
emission in five elliptical galaxies and derived 
some dynamical timescales for gas disks  which are
between 0.07 and 0.2 Gyr.

We first note the presence of peaks in the evolution of the SFRs that are 
associated with recent pericentric passages of 
the satellite. 
Indeed, at these specific times,  an important fraction of available gas is
consumed in starbursts due to  the increase of the gas
density during the interaction.
The last peak in the SFRs is due to the
final plunge of the satellite remnant. 
This behavior can be clearly seen  in Fig. {\ref{sfr1}}, where the
evolution of the star formation rate derived from experiment $B_2$ is plotted 
in parallel with the variation of the radial distance $D$ of the bound part
of the satellite remnant (with respect to the host galaxy center). We can
observe that each SFR peak is preceded by a minimum in  $D$. Physically these
features arise  due to tidal forces that induce instabilities in the satellite
disk, resulting in  loss of  angular momentum via dynamical friction and infall
to the centre.

The presence of several peaks in the  SFR evolution seems to be a general
trend in the case of minor merger events (mass ratio $\leq$ 1:4). A similar
evolution is obtained  even while assuming different star formation
efficiencies or different initial orbital configurations 
(see Fig. {\ref{sfr1}}). We also find similar behaviour
for experiments $B_5$ and  $B_6$ (e.g. for 
mass ratio 1:6 and $r_p=$ 4 and 2 kpc respectively). For reasons of clarity,
 we have omitted showing their evolution in Fig. {\ref{sfr1}}.   

However, the result is different for a major merger event. The right panel of Fig. {\ref{sfr2}} 
indicates that only 2 peaks are observed. In fact,
the timescale of the merging process  decreases as the mass ratio of the system
increases.  The time delays between the final plunge and the first
peak are $\sim 1.75$, $\sim 1.25$ and $\sim 1.$ Gyr for progenitor mass ratios of
1:10, 1:6 and 1:4 respectively. This is mainly due to the
fact that the satellite galaxy gradually loses its energy and angular momentum
under the action of dynamical friction, and finally sinks to the center of
the host galaxy.  This dynamical friction operates more slowly for minor merger
events   (see, for instance, Binney \& Tremaine 2008).  
The ages of the individual mini-bursts
are determined  by  the orbit and its dynamical evolution.

We also note that if the star formation efficiency $c_*$ is low,
a larger reservoir of gas remains  at
the time of the final plunge. This can sustain star formation in the
remnant for a longer time as is clearly suggested in Fig. \ref{sfr1}. 
This result confirms both earlier studies (Larson 1974)
and previous work based on merger simulations which show that gas
can settle in a self-gravitating central disk which may feed a central
black hole  (Mihos \& Hernquist, 1996 ; Barnes, 2002).

 To finish, it should be pointed out that any initial
transient starbursts and numerous fluctuations seen in the star formation rates
before the first pericentric distance 
 may result from  non-equilibrium initial
conditions. Indeed, such starbursts can happen at the beginning of the simulation
since all stars are
unaffected by feedback and then, most of them may be eligible to
form stars. The numerous fluctuations seen in SFRs are also
explained by the fact that a short period is needed for 
type II supernovae to be effective in order to regulate the star
formation in the galaxy.

\begin{figure*}
\rotatebox{0}{\includegraphics[width=17cm]{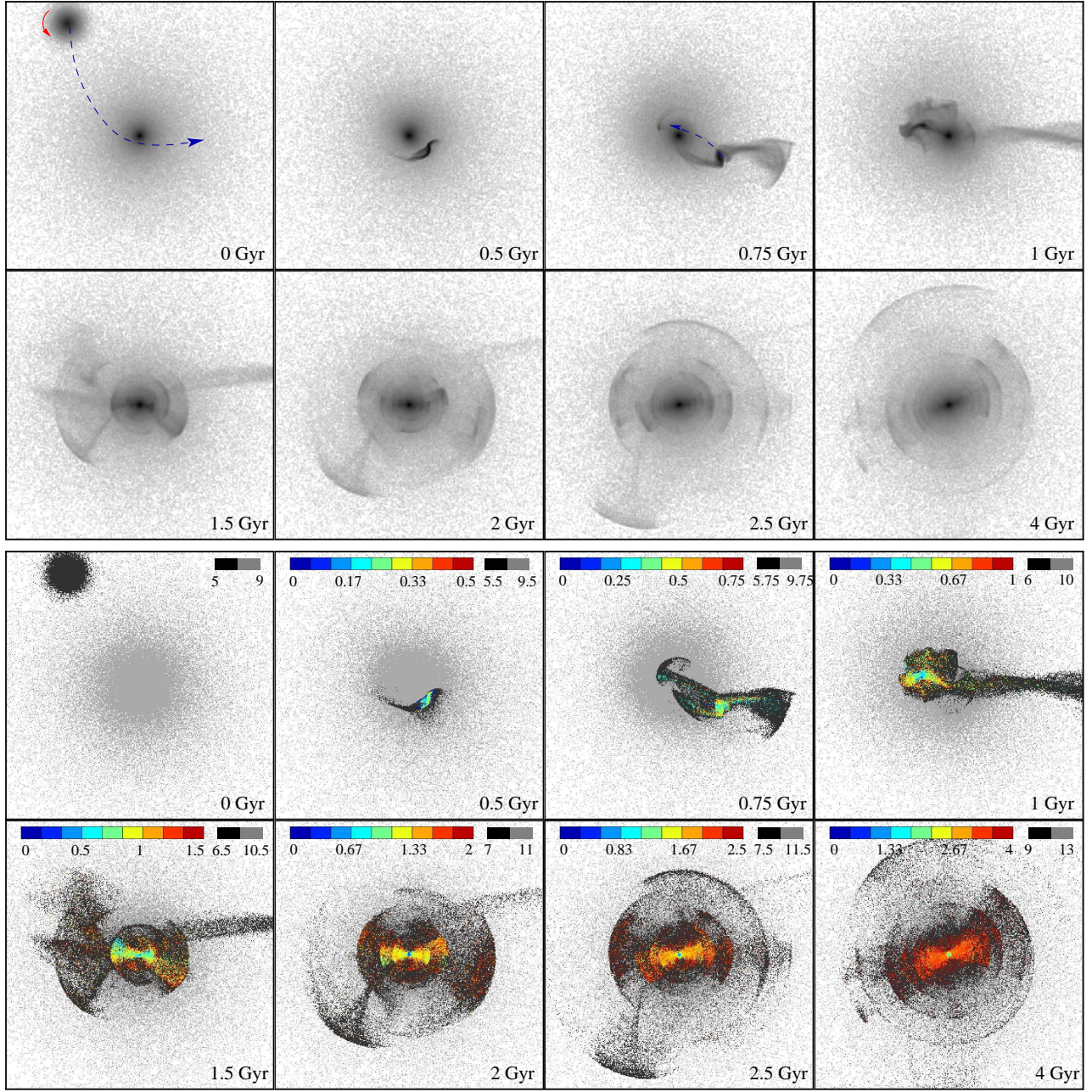}}
\caption{The evolution of the projected spatial distribution of stars
(face-on) from experiment {\it $B_2$}. Lines 1 and 2 show all stars
in grey scale. In the upper panels, the dash blue lines indicate
the trajectories of the satellite while the red arrows shows the initial
spin of the spiral galaxy. Line 3 and 4 represent the stellar ages of
the newly formed stars (in Gyr) at the same specific times. Each square
is $100 \times 100$ kpc in size.}
 \label{stars_distri}
 \end{figure*}

\subsection{Evolution of the spatial distribution of stars}

In this section, we focus on the evolution of the spatial distribution
of stars. In our models,
stars can be divided into 3 distinct groups: those that initially belong to the
elliptical galaxy ($E_*$), those that were present in the disk before the merger
($D_*$), and finally the new stars which have formed during the merger process
($N_*$).
We are particularly interested in the latter stars.
It is worth mentioning that we use the same initial age distribution
as in K09, i.e, we assume that  the elliptical is composed of old
stars  of age $9 $ Gyr whereas the mean age of the stellar population in
the initial disk is   $5 $ Gyr. The motivation of such choices are described
in K09.

\begin{figure}
\rotatebox{0}{\includegraphics[width=\columnwidth]{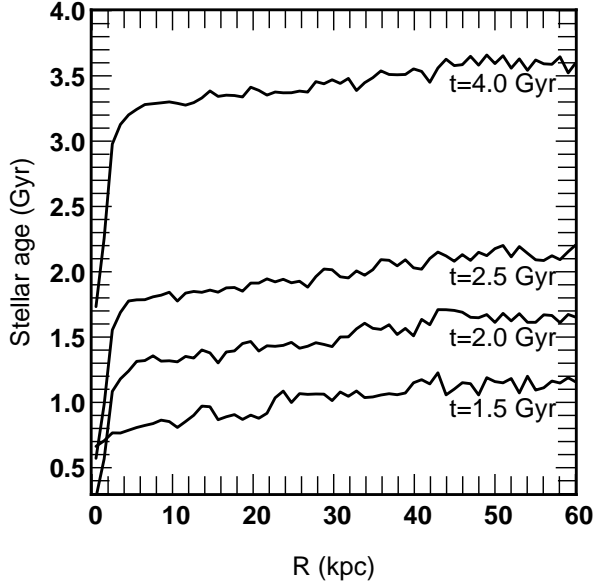}}
\caption{Evolution of the distribution of ages of newly formed stars
(face-on) at different times from the beginning of the simulation for experiment
$B_2$. }
 \label{fig_gradient}
 \end{figure}

Fig. {\ref{stars_distri}} shows the evolution of the projected
distribution (face-on) of all stars in the
experiment $B_2$. After 1.5 Gyr, we notice that the remnant satellite
progressively forms a so-called
``shell galaxy'', in agreement with previous numerical studies
(Quinn 1984; Dupraz \& Combes 1986; Hernquist \& Quinn 1988;
 Hernquist \& Spergel 1992; Hernquist \& Mihos 1995).
The age of all stars is also plotted with a colour coding. 
At large radius, the shell structure
is mainly composed of a fossil stellar population 
of accreted stars $D_*$. 
Such trends are in good agreement with observations of massive ETGs
(see for instance Clemens et al. 2009b). This is mainly due to 
the  dissipationless process and angular momentum conservation
during the merger.  The newly formed stars $N_*$ are
generally formed in the central part of the satellite remnant where the density
is higher. On the contrary,
the gaseous component of the satellite 
falls dissipatively into the potential well of the host galaxy and, after the
final plunge, fuels rapid star formation. 
 This latter phase is clearly active in the inner part of the system
($\leq$ 3 kpc), and after 2 Gyr it may be possible to measure a gradient in the age
of the newly formed stars within $\sim 40$ kpc, as suggested by Fig.
\ref{fig_gradient}.
Since these new stars may have a higher metallicity than the host old ones,
the presence of a metallicity gradient is  also expected,
in agreement  with observations (Clemens et al. 2009b).

\subsection{Size-Evolution of wet merger remnants}

Simple arguments based on the virial theorem allow us to calculate the size evolution
of the collisionless component of the merger remnant (e.g. Bezanson et al. 2009; Naab et 
al. 2009).
Assuming the individual stellar components of the galaxies are in virial equilibrium:
\begin{equation}
E_{i}= - \frac{f}{2} \frac{G M_{i}^2}{r_i} 
\end{equation}  
The factor $f$ depends on the individual functional form of the profile
and is close to $f=0.5$ in the case of de Vaucouleurs and exponential
profiles. The radius $r$ is the half-mass radius of 
the galaxy and $M$ its stellar mass. The merging galaxies are set up
to be on parabolic orbits
with $E_{orb}=0$, which allows us to write down the internal energy of
the remnant assuming energy 
conservation as  
\begin{equation}\label{eq2}
E_{rem}=E_{1}+E_{2}+E_{orb}=-\frac{f}{2} G \left[ \frac{M_{1}^2}{r_1} + \frac{ M_{2}^2}{r_2}  \right].
\end{equation}
Assuming the remnant will settle down in virial equilibrium, then  basic manipulations of Eq. \ref{eq2}
 lead to 
\begin{equation}
\frac{(M_1+M_2)^2}{r_{rem}}= \frac{M_1^2}{r_1}+\frac{M_2^2}{r_2}.
\end{equation}
In the following, we  investigate the merger remnants and characterize
the deviations from this simple virial argument based on the rate of
dissipation and star formation occurring during the merger.
For this purpose, we focus again on experiment $B_2$ which can be 
considered as a typical minor merger scenario. In this case,
we obtain  $r_{rem}\sim 5.67$ kpc.

From the simulation, the half-mass radius $R_e$ at a given time is 
defined to be  the radius of the circle
containing half of the projected stellar mass. 
We have also studied the evolution of effective radius $R_e^V,$ and $R_e^H$
from V and H bands respectively. $R_e^V$ and $R_e^H$ are defined to be the
radius  containing half of the projected flux in each specific band.
In order to take into account the effects of viewing angle,
we have considered 100 different lines of sight chosen randomly and derived
the mean value of $R_e$,  $R_e^V$ and $R_e^H$.

\begin{figure}
\rotatebox{0}{\includegraphics[width=\columnwidth]{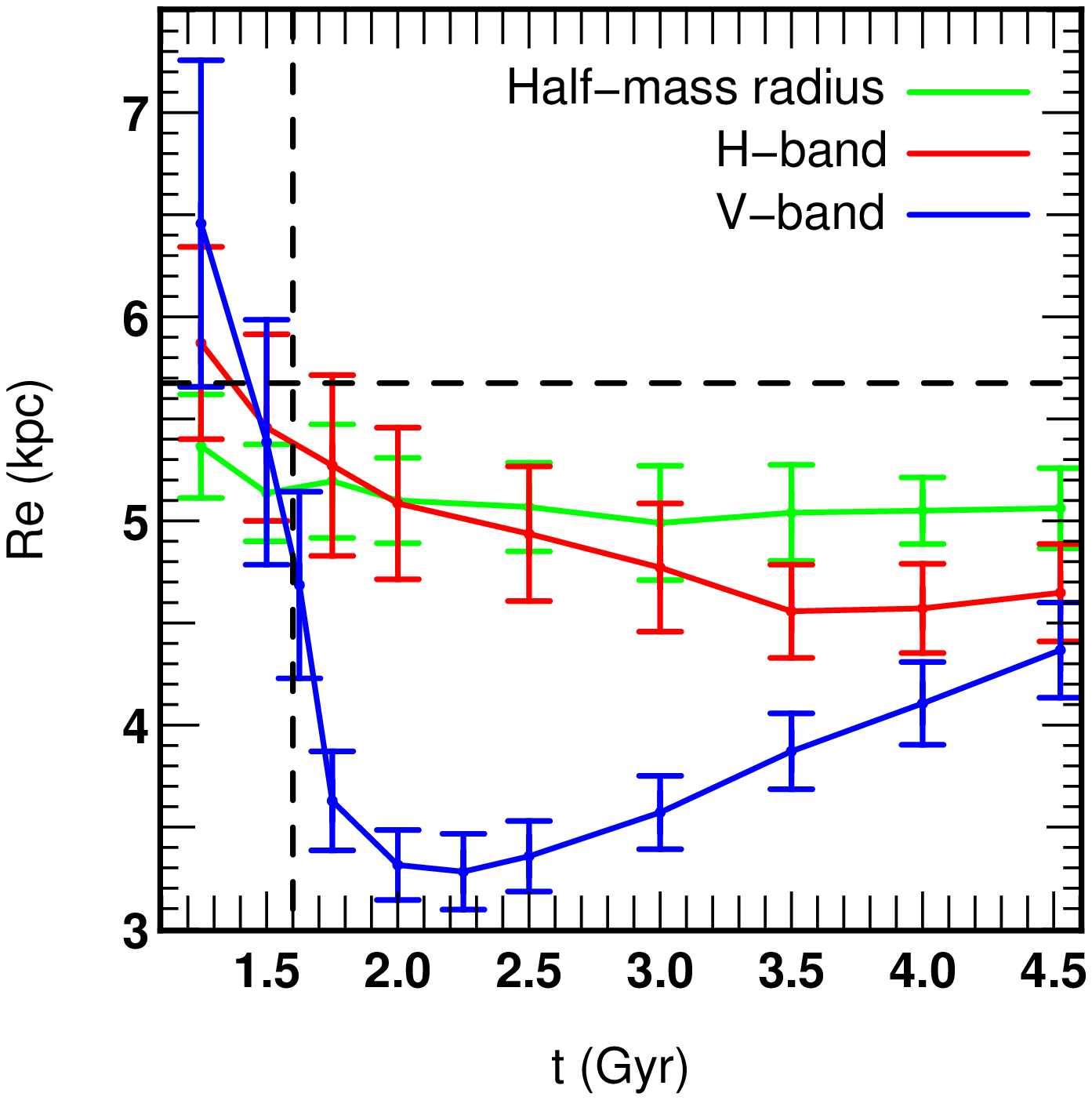}}
\rotatebox{0}{\includegraphics[width=\columnwidth]{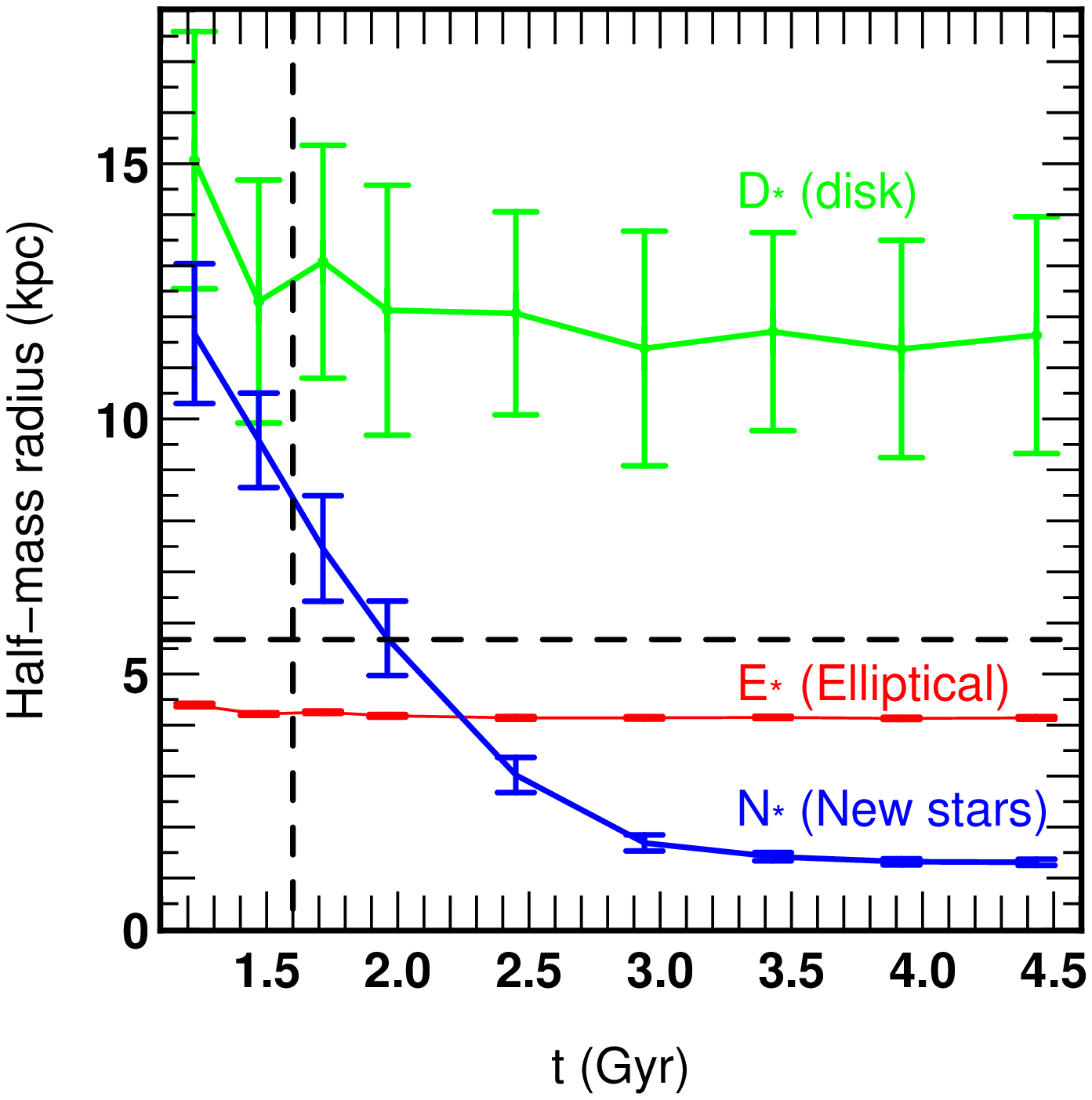}}
\caption{ The upper panel shows the evolution of the half-mass
radius (green line) and effective radius
from V (blue) and H (red) bands for experiment $B_2$.
In the lower panel, we show the evolution of the half-mass radius for each
stellar component: stars that initially belong to the host elliptical galaxy
($E_*$, red line), those that were present in the disk before the merger
($D_*$, green line)
and the new stars which have been formed during the merger process
($N_*$, blue line).
In each panel, the horizontal dashed  line is
the predicted half-mass radius value after the merger from
arguments based on the virial theorem whereas the vertical dashed line is the epoch of the final plunge.}
 \label{fig_size}
 \end{figure}

Fig. \ref{fig_size} shows the evolution of $R_e$, $R_e^V,$ and $R_e^H$ after
 the final plunge for experiment $B_2$. 
The evolution of $R_e$ tends to a constant value ($R_e\sim 5.05$ kpc) which
is lower than the predicted value $r_{rem}\sim 5.67$ kpc.
This result is in agreement with the study of Covington et al. (2007) 
who show that the model systematically over-predicts the radii of
remnants  and using
a wide range of merger simulations which account for dissipative energy 
losses and star formation.

The evolution of $R_e^V$ is different and presents two phases. 
Firstly, a rapid decrease
just after the final
plunge ($t\sim 1.6$ Gyr) which continues for $\sim$ 0.5 Gyr.
This period corresponds to a phase of high
star formation induced by the final plunge as shown in Fig. \ref{sfr1}.
The newly formed stars,
mainly located in the central region,  have a significant contribution 
 of the total flux in the
V-band leading to the reduction in $R_e^V$. In the second phase, beginning
at $t\sim2.25$ Gyr, $R_e^V$ increases due to the fact that the stellar
population becomes progressively older. 
 Indeed, such behavior can be clearly seen in the lower panel of
Fig. \ref{fig_size} where we have plotted the evolution
of the half-mass radius for each stellar component $E_*$, $D_*$
and $N_*$. The half-mass radius of the host elliptical galaxy
tends to a constant value ($\sim 4.2$ kpc) which is actually
very close but lower than 
the initial value ($4.29$ kpc) and probably due to 
the adiabatic contraction induced by the infall of the satellite at the center.
The half-mass radius of stars of the initial disk $D_*$ also tends
to a constant value ( $\sim 11.5$ kpc), which is quite high 
but  explained again by the dissipationless process and angular
 momentum conservation
during the merger. Finally,  the half-mass radius for the newly
formed stars $N_*$ is rapidly decreasing until $t\sim 3.0$ Gyr
due to the high star formation induced after the final plunge. 
After this time, the increase observed in the evolution of  $R_e^V$ 
results from the ageing of the stellar population $N_*$.

The evolution of $R_e^H$   follows the same trend, though the decrease after
the final plunge is less pronounced but continues for a longer time
($\sim$ 2.0 Gyr). The value of  $R_e^H$ is close to $R_e$ values but
still smaller. It is expected that both $R_e^V$ and  $R_e^H$
asymptotically tend to the value of $R_e$ at late time.   
We infer that  because of even modest amounts of star formation,
the half-mass radius differs significantly from  the half-light
radius  especially in the V-band. The observed half-light radius
is expected to  be smaller than the half-mass radius.

\subsection{Evolution of WFC3 bands}

In order to produce synthetic images, we have assumed our galaxy
system to be in the local universe. Such a choice is motivated
by the fact that up to two-thirds of nearby ETGs contains shells
and ripples (see for instance Malin \& Carter 1983; van Dokkum 2005)
and these fine structures and disturbed morphologies  often
coincide with signatures of recent star formation (e.g. Schweizer
et al. 1990; Schweizer \& Seitzer 1992). It is then particularly
interesting to try to characterize this spatially resolved fine structure
whose detection is possible only in our local neighbourhood.
 For instance, using N-body simulations, 
Feldmann, Mayer \& Carollo (2008) have shown that this
broad tidal features cannot be reproduced by equal-mass dissipationless
merger but is well explained by  the accretion of disk-dominated galaxies.
Moreover, minor mergers are expected to be more common 
than major mergers at low redshifts (Khochfar \& Silk 2006;
Genel et al. 2008; Khochfar \& Silk 2009) and
such statements  support our choice of merger mass ratio values.

We start our investigation with  experiment {\it $B_2$}. The 
galaxy pair is supposed to be at a redshift $z\sim 0.023$ (or equivalently
at a luminosity distance of $D=100$ Mpc) in order to facilitate comparisons
with future observational data. In the following, the magnitude units
given as measures of spectral flux are  $W.m^{-2}$,  and in all synthetic
images, we use a logarithmic scale. 

\begin{figure*}
\rotatebox{0}{\includegraphics[width=16.5cm]{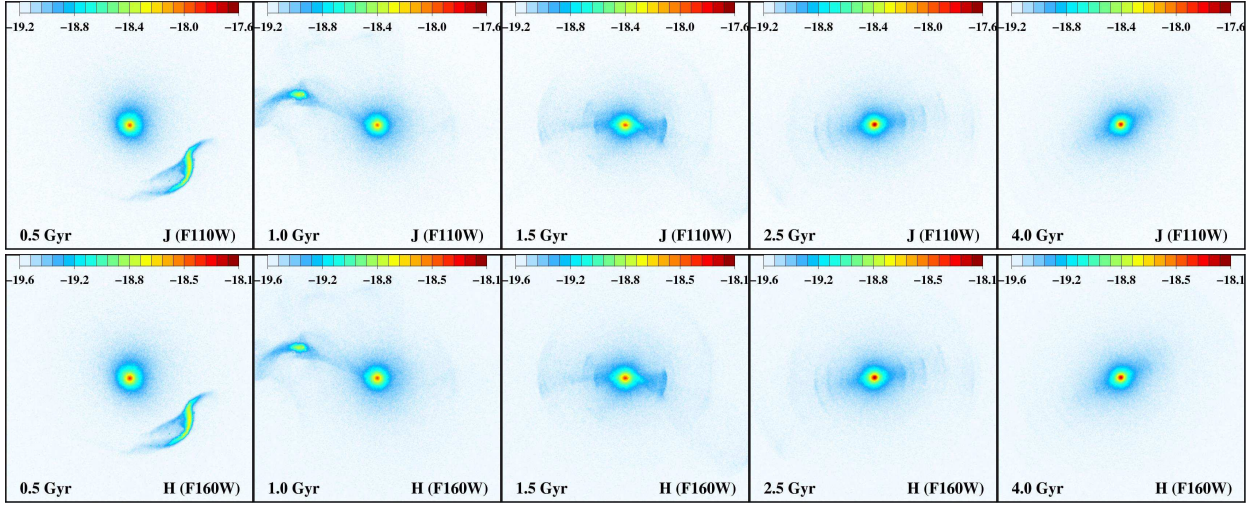}}
\caption{The evolution of synthetic images from experiment
{\it $B_2$} through J (first line)
and H band (second line) assuming the observed system at $z\sim 0.023$.
The images have been Gaussian smoothed with a sigma of 2 pixels.}

 \label{ir_100_6}
 \end{figure*}

\begin{figure*}
\rotatebox{0}{\includegraphics[width=16.5cm]{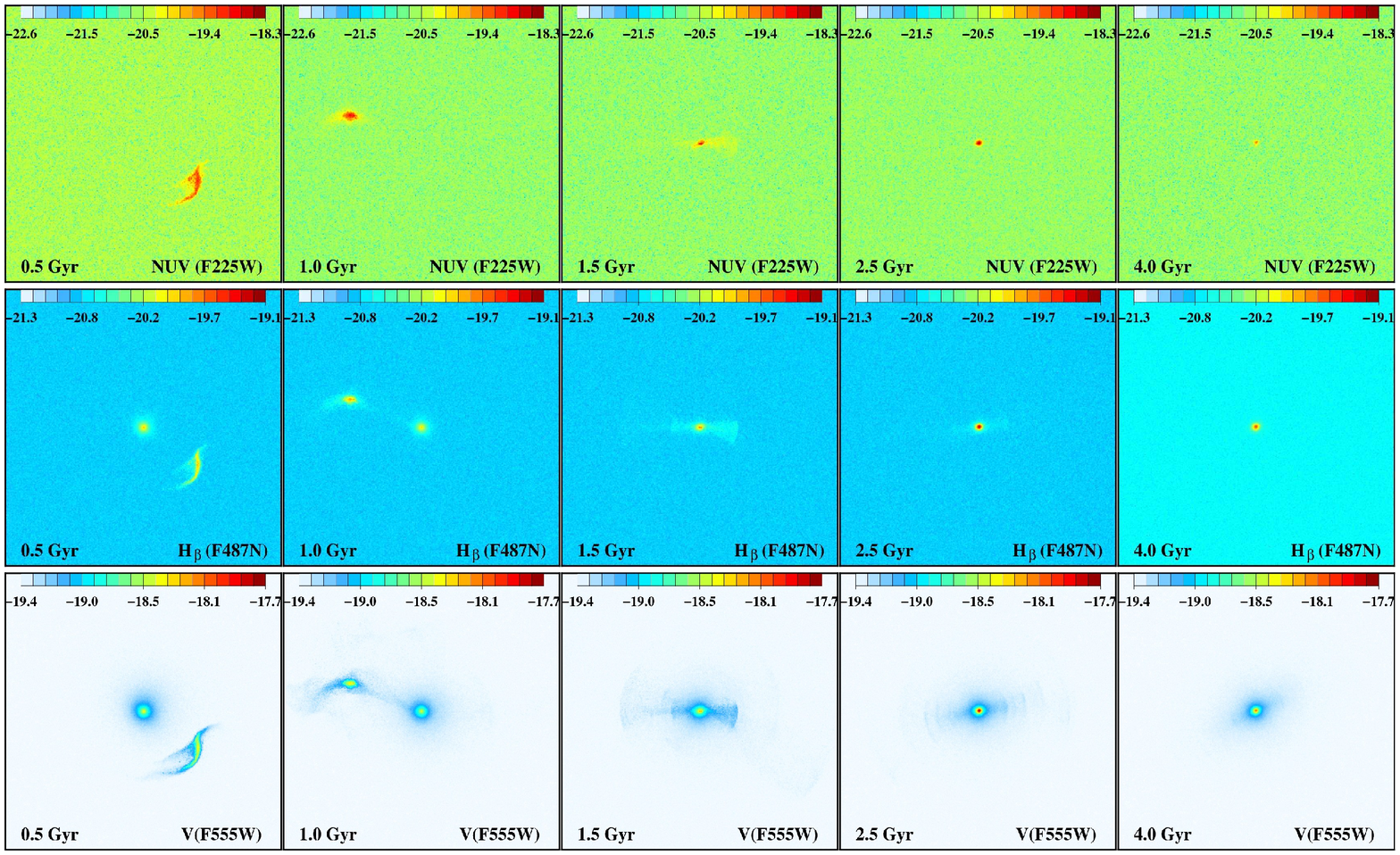}}
\caption{The evolution of synthetic images from experiment {\it $B_2$} through
NUV (first line), H$_\beta$ (second line) and V (third line) band
assuming the observed system at $z\sim 0.023$. }
 \label{uv_100_6}
 \end{figure*}

In Fig. {\ref{ir_100_6}}, we present the grid map derived by our
numerical modelling through the J and H bands.  
It appears that it  is possible to resolve the host elliptical
galaxy and the satellite remnant, in particular the shell structure
which is mainly composed of old stars. 
However, this latter tends to disappear at $t=4.0$ Gyr.
We also note that there are no significant differences  between J and H
bands.  Fig. {\ref{uv_100_6}} shows the evolution of derived synthetic images
from NUV, H$_\beta$ and V bands. The ongoing star formation regions can
be clearly followed through the three filters.
In particular, after the final plunge, ongoing star formation
located at the center of the galaxy can be clearly observed. 
In fact, when comparing with 
Fig. {\ref{stars_distri}}, only stars with ages $\leq 0.5$ Gyr can be
identified. 
Thus,   the combination of  IR and UVIS  images allows us  to 
separate different stellar populations and then distinguish the most bound part
of the satellite remnant, composed of young stars, from the host galaxy,
composed of older stars.
This combination also gives useful  clues on the formation of ETGs:
while the shell structure revealed in IR images support a past merger scenario,
evidence or not of the
presence of young stars in  UVIS  images bring additional constraints on the
wetness/dryness of the merger.

\begin{figure*}
\rotatebox{0}{\includegraphics[width=16.5cm]{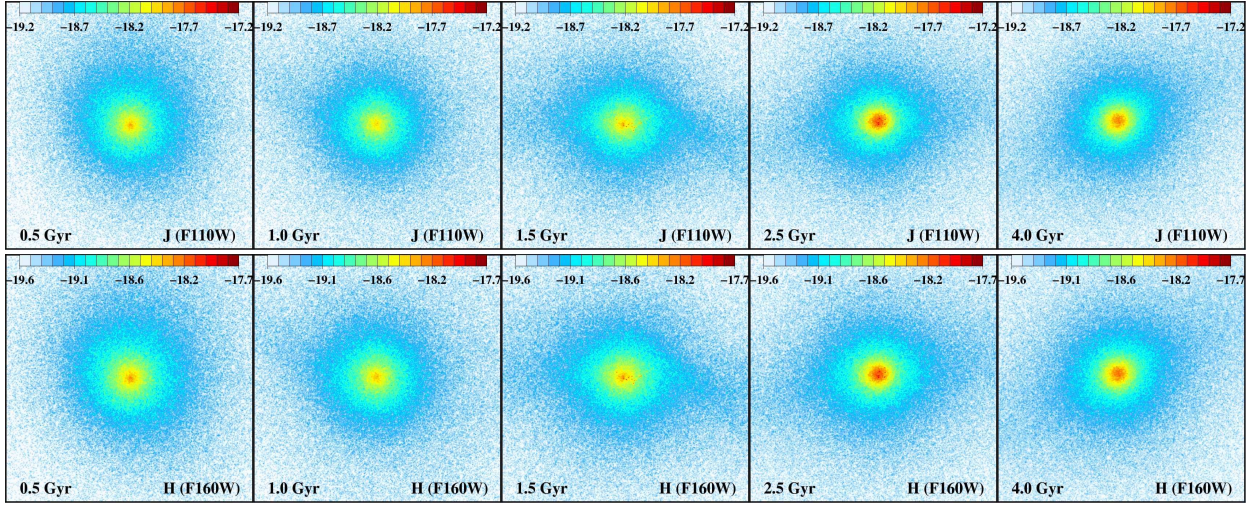}}
\caption{The evolution of synthetic images from experiment
{\it $B_2$} through J (first line) and H band (second line)
assuming the observed system at $z\sim 0.005$. }
 \label{ir_20_6}
 \end{figure*}

\begin{figure*}
\rotatebox{0}{\includegraphics[width=16.5cm]{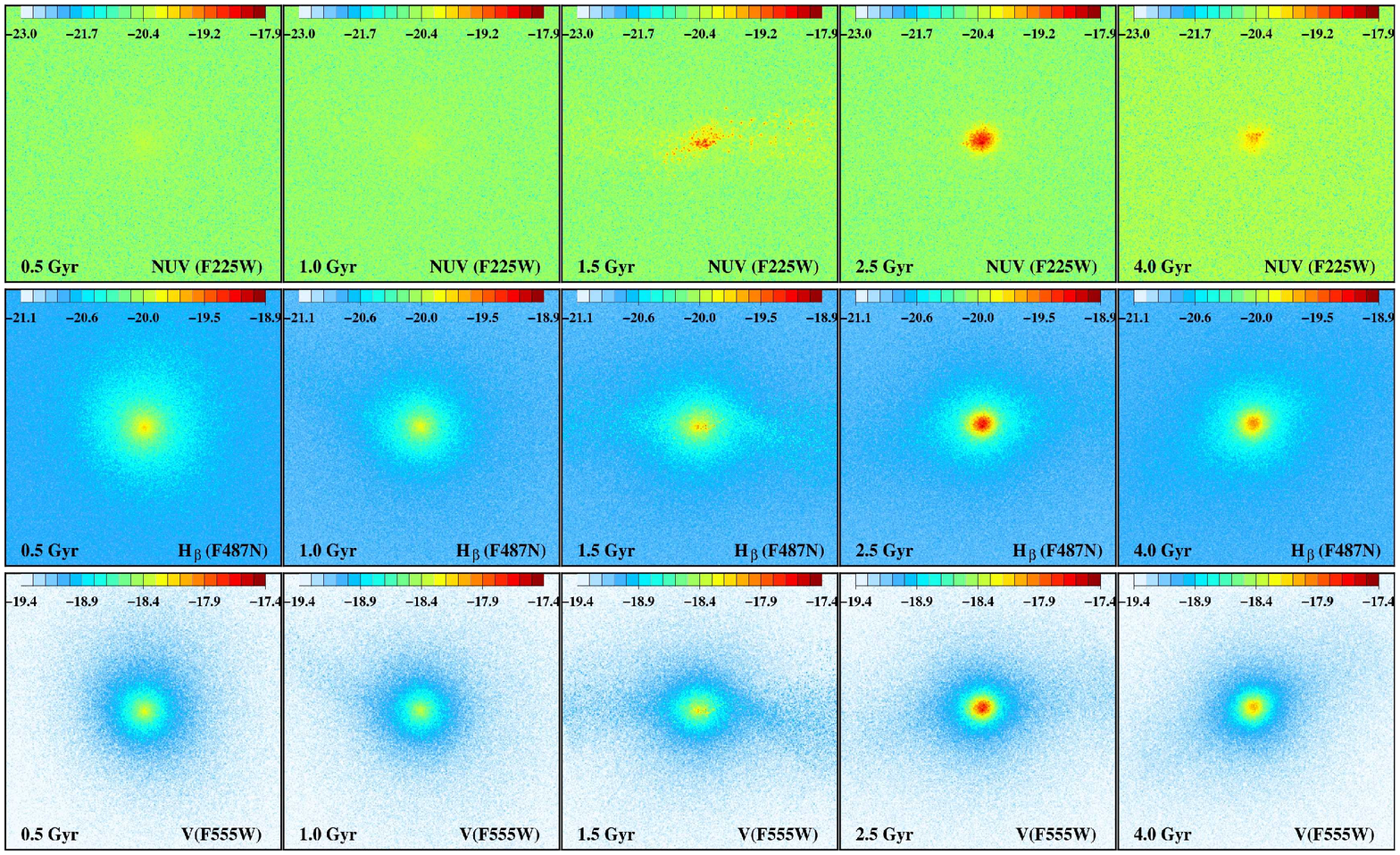}}
\caption{The evolution of synthetic images from experiment {\it $B_2$} through
NUV (first line), H$_\beta$ (second line) and V (third line) band
assuming the observed system at $z\sim 0.005$. }
 \label{uv_20_6}
 \end{figure*}

The same study has been performed for objects assumed to be at 
$z\sim 0.005$ ($D=20$ Mpc). In this case, only the inner part is
imaged due to the instrument FOV, as shown in Figs {\ref{ir_20_6}} and {\ref{uv_20_6}}.
The evidence of recent star formation is again clearly visible 
especially in the NUV band after the final plunge ($t \geq 1.5$ Gyr).
Some structures can also be distinguished around the central part
and represent signatures of a recent merger event.

\begin{figure*}
\rotatebox{0}{\includegraphics[width=16.5cm]{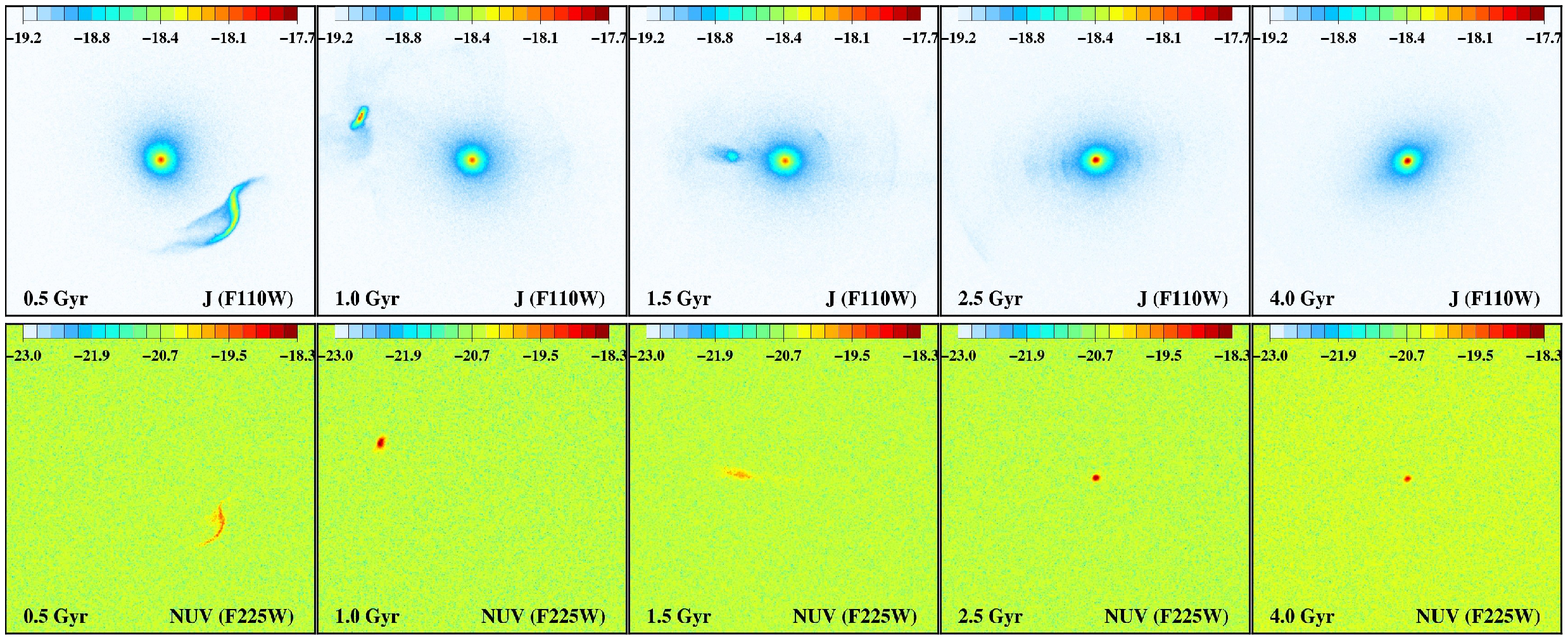}}
\caption{The evolution of synthetic images from experiment {\it $B_1$} 
(mass ratio 1:6 and $c_*$=0.01) through
J (first line) and NUV (second line) assuming the observed system at $z\sim 0.023$.
}
 \label{ir_uv_6}
 \end{figure*}

We have also studied how the free parameters (star formation efficiency and
initial mass ratio) affect our results. 
For instance, Fig. \ref{ir_uv_6} shows the evolution
of synthetic images from the J and NUV bands in the 
experiment {\it $B_1$}. In this case, although the star formation efficiency
is lower, the satellite remnant can also be identified in the NUV band. In
the J band, the shell structure is less pronounced.

\begin{figure*}
\rotatebox{0}{\includegraphics[width=16.5cm]{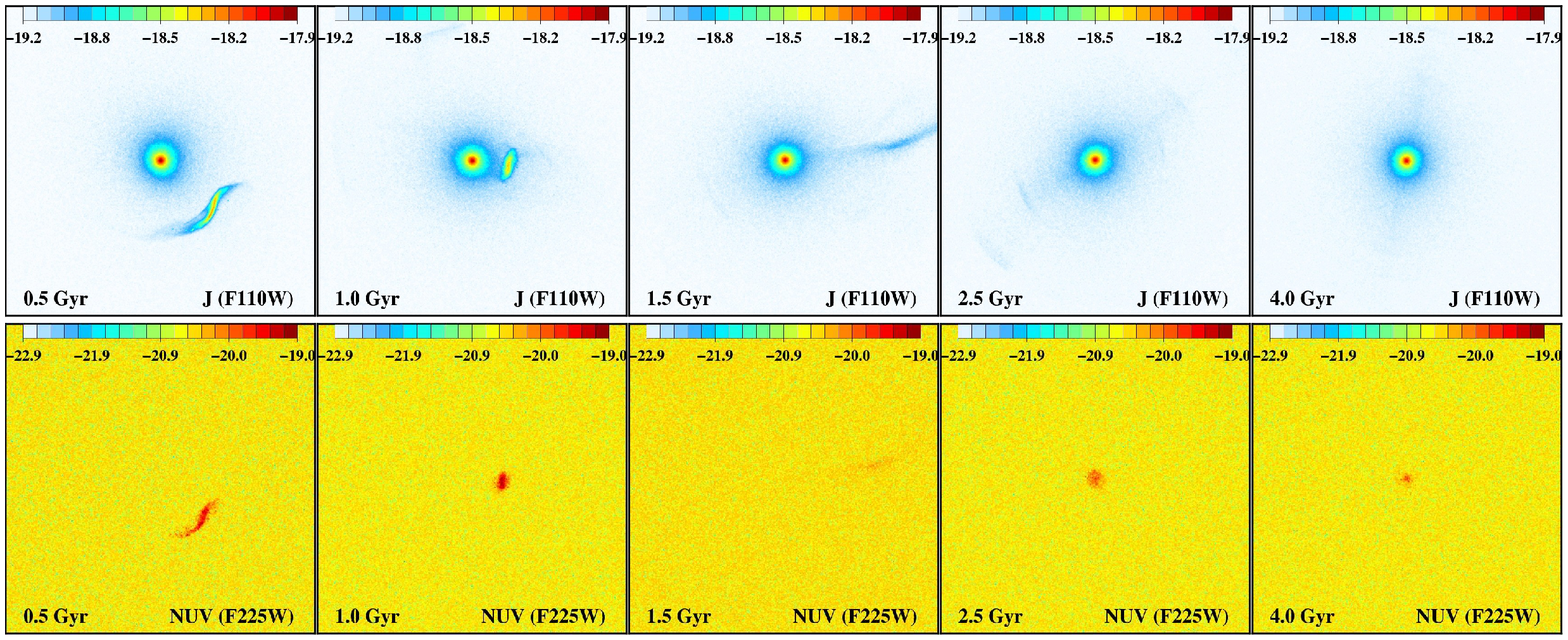}}
\caption{The evolution of synthetic images from experiment {\it $A_2$}
(mass ratio 1:10 and $c_*$=0.05) through
J (first line) and NUV (second line) assuming the observed system at $z\sim 0.023$. 
}
 \label{ir_uv_10}
 \end{figure*}

\begin{figure*}
\rotatebox{0}{\includegraphics[width=16.5cm]{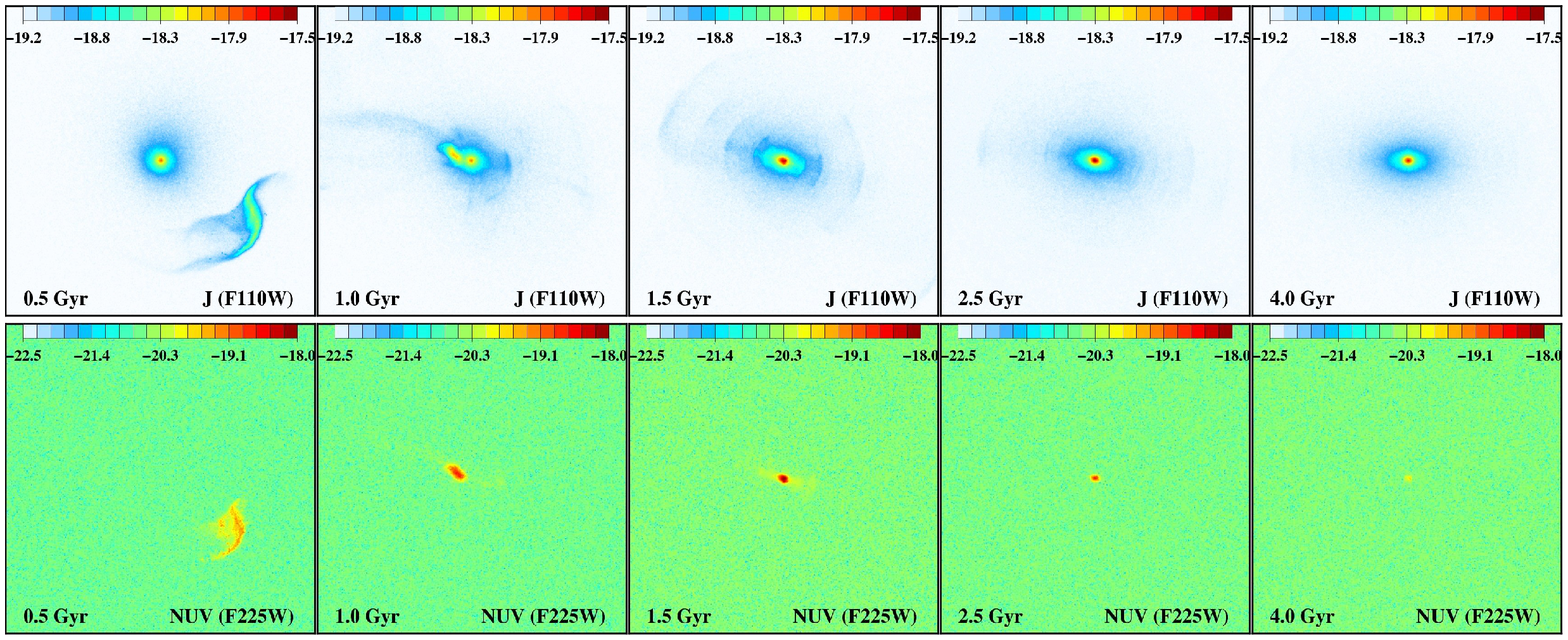}}
\caption{The evolution of synthetic images from experiment {\it $C_2$}
(mass ratio 1:4 and $c_*$=0.05) through
J (first line) and NUV (second line) assuming the observed system at $z\sim 0.023$. 
}
 \label{ir_uv_4}
 \end{figure*}

Comparison with  mass ratios 1:4 and 1:10 presents similar conclusions
although the structures are better resolved in the
case of the 1:4 mass ratio due to the higher 
star formation induced during the merger
(see Fig. \ref{ir_uv_10} and \ref{ir_uv_4}). 
for instance  maximum values in the NUV band between 1:4 and 1:10
   merger mass ratios  differ by up to an order of magnitude while they are
  comparable in the H band.
However, 
simulations with  the 1:4  merger mass ratio should be regarded
as a limiting  case, since it is close to the major merger scenario.

\section{Discussion}

The formation of early-type galaxies remains a major unresolved problem in
astrophysics, and  in particular their star formation history is known to
be more complex than that of a 
single population model. Studies of nearby
galaxy populations can provide useful clues to resolving this problem, by setting
tight constraints on the level of the most recent star formation activity.
In the present paper, we have used idealized hydrodynamical simulations,
including a standard star formation prescription, to study the minor
merger process between an elliptical and a spiral galaxy which is expected to be the
dominant  channel in terms of the morphology of the merging galaxies (Khochfar \& Burkert 2003).
There is now observational evidence that
massive ETGs may have undergone not more than one major merger 
since $z\sim0.7$ (Bell et al. 2006; McIntosh et al. 2008).
Such results are supported 
by theoretical evidence that massive halos ($\ge 10^{11} M_\odot$)
experience only one major merger between $z=2$ and $z=0$,
while minor mergers are much more common (Khochfar \& Silk 2006;
Genel et al. 2008; Khochfar \& Silk 2009).
We use idealised simulations in this work for two main reasons.
Firstly, it allows us to reach a high mass resolution ($\sim 10^{4-5} M_\odot$)
with  reasonable computational costs compared to cosmological simulations.
Secondly, it permits us to consider a wide range of free parameters (including star formation
efficiencies and initial orbital parameters) and to study how the results change
accordingly. 

We have generated 15 experiments by considering  different merger mass ratios,
(1:10, 1:6, 1:4 and 1:1) and three star formation efficiencies ($c_*=$0.01,
0.05 and 0.1).
We find that the satellite is progressively
disrupted by tidal torques and the star formation induced
by the merger process shows several peaks in its evolution, associated
with the passage of the remnant at decreasing pericentric distances. 
The satellite remnant progressively forms a shell structure within the host galaxy, which is
composed of old stars at large radii and is explained by the fact that
the stellar accretion and ensuing relaxation
occurs via dissipationless mechanisms. On the contrary,  the center is
dominated by the newly formed stars due to the infall of the gas   
to the centre in a dissipative process. A positive radial age 
gradient is clearly seen and remains constant in time.
All of these results are consistent with the conclusions
of observational studies of massive ETGs (see for instance, Clemens et al. 2009b).

These typical shell structures are of particular interest as a means of  
 characterizing the last merger experienced by the 
host galaxy, i.e, the mass ratio, dryness or orbital configuration.
For instance, recently Canalizo et al. (2007) have tried to set constraints
on the age of the plausible merger responsible for  shell structures observed
in a quasar host galaxy from HST/ACS. By using N-body simulations, they found that
the merger occurred a few hundred Myr to $\sim$ 2 Gyr ago when matching the
observed and simulated outermost shell and causal connection with the quasar activity.
 But the  accurate estimation  of the age of the merger  proves to be a very
hard task due to the
high number of orbital parameters,  even when combining dynamical and
photometric properties of the galaxy. For instance,
Di Matteo et al. (2009) investigated the evolution of star formation efficiency
in galaxy interactions and mergers using numerical simulations 
and found that  retrograde orbits or high pericentric distances
favor the enhancement of the star formation.  Such degeneracy may be
difficult to deal with. 

Our main result is to demonstrate that colour information
provides an important additional constraint via the induced star formation that
together with the associated morphology provides further  and important insights
into the merger parameters. 
In particular, future observations especially with the resolution and NUV power
of WFC3 will open new vistas on star formation histories of ETGs. We have  derived
realistic synthetic images through J, H, NUV, H$_\beta$ and V filters  that can be
compared with  future WFC3 observations. 
Comparison of synthetic maps in the NUV and J bands for different merger
mass ratios and star formation efficiencies gives similar results:
evidence of recent star formation can be clearly detected before and
after the final plunge,  especially in the NUV band.

Our study does not include AGN feedback. We will consider this in a
future paper. Here we note that while such a mechanism is supposed to play
an important role in the early phases of the spheroid star formation,
its role in the late phase is still poorly known.
As suggested by our results,  the infall of the gas at the final plunge
can settle in a self-gravitating central disk which may feed a central
black hole.  For instance, N-body simulations including 
feedback from black hole accretion
indicate that the quasar activity is coincident with the coalescence
of the nuclei and eliminates all star formation afterwards
(Springel, Di Matteo \& Hernquist 2005).
This will probably lead to a reduction
of the total flux in the UVIS bands.

In summary,  minor mergers seem to play a fundamental role in the late phases
of ETG evolution. Minors mergers may be the
principal mechanism behind the large UV scatter and associated low-level
recent star formation observed in ETGs in the nearby universe
(Kaviraj et al. 2009). In general, we find composite populations
contributing to the star formation rate which persists for 2-3 Gyr,
corresponding to several orbital times. Minor mergers naturally lead
to star formation that is extended over a much longer period than for
the single population starburst.  The limiting case 
of major mergers results in much shorter star formation time-scales.
Size evolution is another natural consequence of the minor merger model.
The considerable amount of dissipation and central star formation reduces
the apparent half-light radius of the merger remnant compared to its
half-mass radius (see Fig. 6), thus indicating that size determinations
based on rest-band magnitudes sensitive to star formation will clearly
underestimate sizes by factors of 1.2-1.6 for a few Gyr after the
merger. Taking our results at face value and considering that the
frequency of gas-rich minor mergers is larger at earlier times, the
size-evolution of elliptical galaxies from high to low redshift could
be milder than assumed so far.

To conclude, the present work shows that 
minor mergers induce amounts of star formation in ETGs
which can be measured through UV bands. 
WFC3 represents the best instrument
to study these minor merger events because it has a matching UV and optical
FOV and gives the resolution to see young 
substructures (which would not be possible with GALEX for instance).
The previous ACS/HRC UV detectors had a
tiny FOV so it was not possible to study a galaxy up to 1 effective radius
because one would
get the whole galaxy in the optical image and only a fraction of
its core using the UV. With WFC3, it is now possible to map the entire galaxy
in both the UV and optical making it possible for the first time
to perform spatially resolved star formation histories in ETGs
at low redshift using the UV. 
Moreover the ability to see the young substructures is important
because it rules out UV flux from old sources such as horizontal
branch stars (which would follow the optical light profile).

\vspace{1.0cm}

\noindent
{\bf Acknowledgment}

\noindent
S.\,P. acknowledges support from  ``l'Agence Nationale de la Recherche'',
R.M.\,C. and S.\,G. from the STFC, and S.\,K. through an Imperial College
Junior Research Fellowship, a Research Fellowship from the Royal
Commission for the Exhibition of 1851 and a Senior Research
Fellowship at Worcester College, University of Oxford.
We warmly thank the referee for his useful comments that helped to improve
the text of this paper.


\end{document}